\documentclass[useAMS,usenatbib]{mn2e}

\usepackage{txfonts}
\usepackage{subfigure, epsfig}
\usepackage{amssymb}
\usepackage{amsfonts}

\usepackage{mncite}

\newcommand{\ie}{i.e.}

\newcommand{\cmeter}{{\,{\rm cm}}}

\newcommand{\gram}{{\,{\rm g}}} 
\newcommand{\msol}{{\,{\rm M}_\odot}} 
 
\newcommand{\ex}[1]{\cdot 10^{#1}}

\author[M. Britsch, C. J. Clarke and G. Lodato]{M. Britsch$^{1,2,4}$, C. J. Clarke$^{1}$ and G. Lodato$^{3,1}$\\
$^1$Institute of Astronomy, Madingley Road, Cambridge, CB3 0HA \\
$^2$ITA, Zentrum f\"ur Astronomie der Universit\"at Heidelberg, Albert-Ueberle-S
tr. 2, 69120 Heidelberg\\
$^3$ Department of Physics and Astronomy, University of Leicester, Leicester, LE1 7RH \\
$^4$ Present address:  Max-Planck-Institut f\"ur Kernphysik, Saupfercheckweg 1, 69117
Heidelberg, Germany
}

\title{Eccentricity growth of planetesimals in a self-gravitating protoplanetary
 disc}

\begin{document}

\maketitle
\begin{abstract}
We investigate the orbital evolution of planetesimals in a self-gravitating
circumstellar disc in the size regime ($\sim 1-5000$ km) where the 
planetesimals behave approximately as test particles in the disc's
non-axisymmetric potential. We find that the particles respond to the
stochastic, regenerative spiral features in the disc by executing
large random excursions (up to a factor of two in radius in $\sim
1000$ years), although typical random orbital velocities are
of order one tenth of the Keplerian speed. The limited time frame and
small number of planetesimals modeled does not permit us to discern any
{\it net} direction of planetesimal migration. Our chief conclusion is that
the high eccentricities  ($\sim 0.1$)
induced by interaction with spiral features
in the  
disc is likely to be highly unfavourable to the collisional growth of
planetesimals in this size range  while the disc is in the self-gravitating regime.  
Thus {\it if}, as recently argued by Rice et al 2004, 2006, 
the production of planetesimals gets under way when the disc is in the
self-gravitating regime (either at smaller planetesimal size
scales, where gas drag is
important, or via gravitational fragmentation of the solid component), then
the planetesimals thus produced would not be able to grow collisionally
until the disc ceased to be self-gravitating. It is unclear, however,
given the large amplitude excursions undergone by planetesimals in the
self-gravitating disc, whether they would be retained in the disc throughout
this period, or whether they would instead be lost to the central star.

\end{abstract}


\section{Introduction}

It is currently unclear whether the process of planet formation 
begins during the earliest
phases of star formation (i.e. in circumstellar discs that are
strongly self-gravitating). The high disc mass at this stage
(with gas mass $\sim 10 \%$ of the central star's mass 
(e.g. Eisner and Carpenter 2006) is favourable
to this hypothesis and provides the only opportunity for gas 
giants to form via (gas phase) gravitational instability \citep{boss00}. A more
subtle effect - again strongly favoured during the self-gravitating case - arises
through the interplay between pressure gradients in spiral arms and
gas drag on solid particles and results in the accumulation of dust
in spiral arms \citep{rice04}. This dust may then be able to accumulate
further through the action of collisions and/or self-gravity in the dust
phase \citep{rice06}. \footnote{Note that such rapid accumulation, which might result in the rapid formation of large size planetesimals, is required in order for the growing planetesimals to escape the dangerous meter-size range, where they are subject to very fast migration down into the hottest inner disc \citep{weiden77}.}  

 Once the disc is no longer self-gravitating in the dominant (gas)
component, it is evidently impossible to form planets through either gas phase
Jeans instability or dust accumulation in spiral arms and thus the most
successful models invoke the collisional growth of dust \citep{pollack96}
followed possibly by the accretion of a gaseous envelope. Despite the lower
surface densities during later phases (estimates of gas disc mass from
sub-millimetre dust measurement suggest typical masses that are around an
order of magnitude lower than in the self-gravitating case; \citealt{andrews05}), the great advantage for planet formation is simply the
longer duration of this later phase. Discs with sufficient gas and dust to form
planets typically survive for $10^6-10^7$ years \citep{haisch01,armitage03,wyatt05}, whereas the existence of dusty
debris discs  around older stars suggests that (potentially planet building)
collisions between rocky planetesimals proceeds for $10^8$ years or
more \citep{bromley04}. These numbers should be contrasted with the brief
window of opportunity ($\sim 10^5$ years) in the self-gravitating phase.

 In summary, then, it is unclear without detailed calculation which of these
regimes is favoured in the trade-off between lifetime and disc mass.

 The viability of planet formation in the self-gravitating regime however
requires that a number of conditions are satisfied beyond the initial
phases of accumulation outlined above. In particular, we need to understand
the dynamical evolution of the protoplanetary components (whether at the
scale of dust aggregates, planetesimals or proto-giant planets) with
a view to answering the following questions: (i) does orbital migration
result in such bodies spiralling into the central star, or can they
be retained in the disc at least over the initial self-gravitating
phase, and (ii) (in the case of dust or planetesimals) are the kinematics
of such bodies conducive to further collisional growth?

 As a first step in answering this question we here undertake pilot simulations
of a small number of test particles in a self-gravitating disc, with a
view to both considering the issue of their orbital migration and  to
establishing the velocity dispersion that is set up through interaction
with spiral structure in the disc. It should be stressed that the calculation
differs in several important  ways from most calculations of orbital migration
and planetesimal kinematics  reported in the literature. Firstly,
conventional planetary migration, whether classified as Type I or Type II, is
the consequence of the gravitational torque exerted on the planet 
(or planetesimal)  by
non-axisymmetric structure {\it induced by the planet/planetesimals}. 
Since this torque
contribution can have significant contributions at the size scale
of the Hill radius (or below: \citealt{bate03}), it is evidently important
that this scale is well resolved in numerical calculations, which makes
the modeling of the migration of low mass objects particularly challenging, 
especially with codes such as Smoothed Particles Hydrodynamics (SPH) 
\citep{valborro06}. 
In the present case, however, we are interested in  a regime where the
disc has pronounced non-axisymmetric structure {\it in the absence of
the planet/planetesimal} 
(due to the presence of self-gravitating spiral modes in the
disc) and we are therefore not obliged to resolve the Hill radius in
order to capture the dominant torque contribution.  Evidently, as
the mass of the planetesimal is increased, the contribution to
the torque which it experiences from self-induced spiral structure
becomes significant and we will assess this {\it a posteriori} in order
to set an upper mass scale on the applicability of our
calculations. Secondly, in
conventional calculations of the kinematics of planetesimal swarms
\citep{kokubo2000,thommes03}, the velocity dispersion of planetesimals
is set by an equilibrium between what is sometimes termed 'viscous
stirring' (i.e. the transfer of kinetic energy from larger bodies
to the planetesimal swarm via two-body scattering) and eccentricity
damping by gas drag. In the present case, however, the velocity dispersion
of our  particles is set by the level to which eccentricity is
pumped through interaction with the fluctuating potential of the
self-gravitating disc. In this pilot calculation we omit gas drag, which
places a lower limit to  the size scale of objects to which the
calculation is applicable; again, we set this lower limit {\it a posteriori}.

  Although we have stressed the qualitative differences between this
calculation and that reported in the large body of literature on
planetary migration and planetesimal kinematics, we point out that it
bears close comparison with the calculation of \citep{nelson05}, which
studies the response of planetesimals to the fluctuating non-axisymmetric
structure in a disc that is subject to the magneto-rotational
(MRI) instability. In this study it was found that the planetesimals
underwent stochastic migration, but that it was not possible to
discern the mean magnitude (or even sign) of the migration rate over the
limited timeframe of the simulation (a few hundred orbits). In our
experiment also, conducted over a similar time frame to Nelson (2005),
we are unable to discern a net direction of migration, but we find
(perhaps unsurprisingly given the fact that our disc is about an
order of magnitude larger in mass) that the amplitude of the stochastic
variations in the orbital radius is much larger, with several planetesimals
undergoing radial excursions of a factor two or more in both directions.
Evidently, this behaviour has important implications for the ability
of planetesimals to grow by collisions during the self-gravitating phase.

  The structure of the paper is as follows. In Section 2 we describe
the modeling of the self-gravitating disc, which is maintained in
a 'self-regulated' (i.e. non fragmenting) state through the imposition
of an {\it ad hoc} cooling law for which the cooling time is a suitably
large multiple of the local dynamical time. In Section 3  we analyse the
orbital response of $12$ test particles placed in a disc 
of mass  $0.1$ and $0.5 M_\odot$.
We also analyse the location of the dominant torque contribution
on the planetesimals and assess the range of planetesimal mass and size
scales for which we expect 'test particle' behaviour and for which
gas drag can be ignored. In section 4 we discuss the implications of
our numerical findings for the migration and collisional growth
of planetesimals in self-gravitating discs. Section 5 contains
our chief conclusions.

\section{Numerical simulations}

\subsection{Numerical code}
\label{s:numerical}

In this paper we simulate the interaction of a gaseous self-gravitating disc with 
a small number of planetesimals using SPH, a Lagrangian particle based method \citep{lucy77,benz90,monaghan92}. This allows the simulation of gaseous and $N$-body 
particles, using individual time-steps \citep{bate95}, thus resulting  in a large saving in 
computational time when a large dynamic range of timesteps is involved. 

The general setup of our simulations is similar to \citet{LR04,LR05}. We consider a massive
gaseous disc orbiting around a $1M_{\odot}$ central star. The disc is taken to extend initially
from 0.25 to 25 AU and we have considered two cases, where the disc mass is equal to $0.1M_{\odot}$ (standard runs) and $0.5 M_{\odot}$ (massive runs), respectively. 

Our simulations employ a number $N=250,000$ particles and are thus intermediate-high resolution simulations. Previous simulations \citep{LR04,LR05} at the same resolution have been shown to reproduce reasonably well  the internal disc dynamics. In particular, once the disc reaches a quasi-steady self-regulated state (see below), the  disc thickness $H$ is resolved with an average of 2 smoothing lengths for the low disc mass case and with 4 smoothing lengths for the hotter, high disc mass case. 

As mentioned above, our code follows the dynamics of both SPH particles and of point masses, which are able to accrete gas particles, if they come closer than a given accretion radius to the point mass. For the central object, such accretion radius is set to $0.25$ AU. At the end of the simulations typically a fraction no larger than 10\% of the initial disc mass has been accreted onto the
central star. The planetesimals have a nominal mass equal to the default gas particle mass, that is $4\times 10^{-7}M_{\odot}$ for the standard runs and $2\times 10^{-6}M_{\odot}$ for the massive run. This choice ensures that the gravitational force of individual planetesimals does not influence the overall disc structure and the dynamics of the planetesimals. In order to prevent planetesimals from acceting gaseous particles, we have set their accretion radius to a very small value. 

The gas particles are allowed to heat up via $p\mbox{d}V$ work and artificial viscosity and  to cool down with the following simple prescription 
\begin{equation}
\left.\frac{\mbox{d}u}{\mbox{d}t}\right |_{\rm cool} = - \frac{u}{t_{\rm cool}}
\end{equation}
where the cooling time-scale $t_{\rm cool}=\beta\Omega^{-1}$
and $\beta $ is fixed in space and time. While this is
not a realistic prescription for the cooling in  actual protostellar discs,
it provides a convenient way of parameterising the cooling physics in order
that one can ensure that one is modeling discs that remain in the
self-gravitating regime without fragmenting.  
 In particular, if the value of the parameter $\beta$ is large enough ($\beta\gtrsim 5-6$; \citealt{gammie01,RLA05}), it allows the realization of a feedback loop where the disc evolves into a quasi-steady, self-regulated case, and is maintained close to marginal stability. For smaller values of $\beta$ the disc undergoes fragmentation. In our simulations we have set $\beta=7.5$.  
We note that since this is close to the critical value of $\beta $, we expect
such a disc to be relatively `active' (in the sense that the maintenance
of thermal equilibrium requires spiral modes to be of relatively large
amplitude). We revisit this point in Section 4.

  \subsection{Initial conditions}

Initially the gas in the disc is relatively hot, with a temperature profile $T(R)\propto R^{-1/2}$ and a minimum value of the stability parameter \citep{toomre64} $Q=2$, attained at the outer disc edge. Since marginal stability corresponds to $Q=1$,  the whole disc is initially gravitationally stable. We initially follow the evolution of the gaseous disc only, as it cools down and becomes gravitationally unstable. The initial evolution is very similar to the one described in several other papers for example \citep{LR04,LR05,mejia05}: when $Q\approx 1$, the disc develops a spiral structure, inducing dissipation through compression and mild shocks and therefore heating up the disc and establishing a feedback loop which keeps the disc close to marginal stability, so that  the Toomre $Q$ parameter
is close to unity over the radial range $\sim 1-20$. This initial evolution takes roughly  $60$ outer
dynamical times  for the standard runs and $45$ outer dynamical times for
the massive runs. In this state, the surface density distribution is dominated by pronounced
(of order unity in amplitude) spiral features, which, although regenerative,
are individually transient structures (i.e. with lifetimes that are of order
the local dynamical timescale). 

\begin{figure}  \centering  %
\includegraphics[width = 0.45\textwidth]{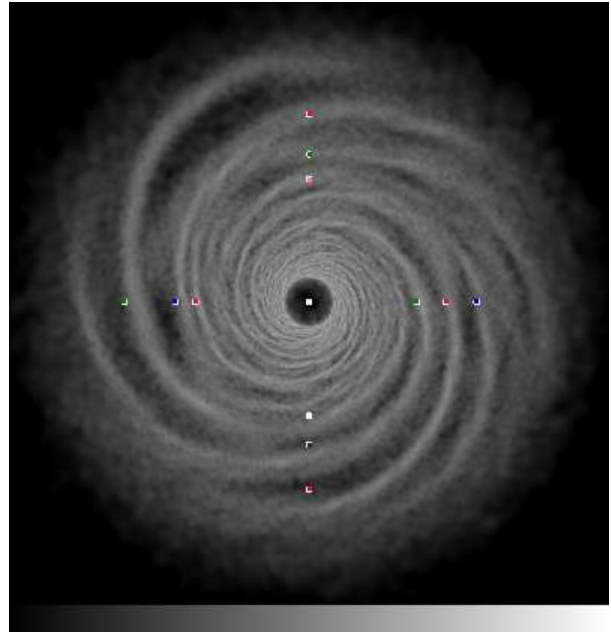}  
\caption[The planetesimal position at the beginning]{
Initial positions of the planetesimals in the evolved quasi-steady spiral structure of the 0.1 $\msol$ gas disc at the beginning of the run. The dimensions are 60 AU in each direction and the logarithmic colour scale shows the surface density $\Sigma$ and goes from 10 to $10^{5} \gram \cmeter^{-2}$.}  
\label{f:planetpos}
\end{figure}

\begin{figure}  \centering  
\includegraphics[width = 0.45\textwidth]{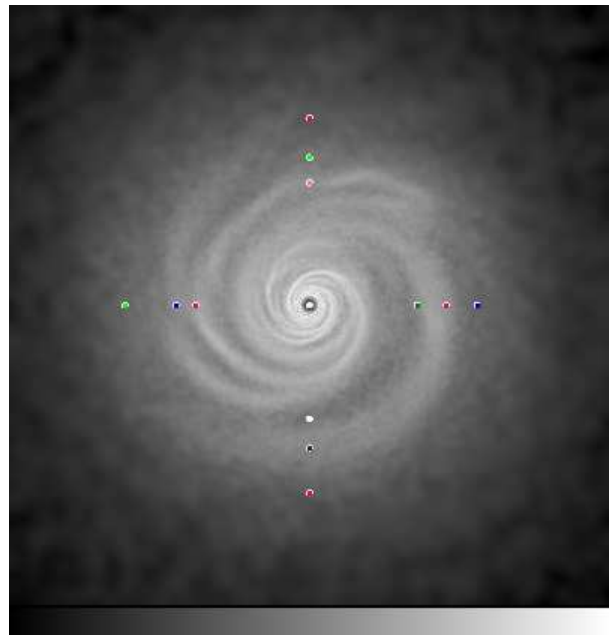}  
\caption[The planetesimal position at the beginning high mass run]{
Same as Fig \ref{f:planetpos} but for the 0.5 $\msol$ gas disc. Note the warmer conditions
in the massive disc in the self-regulated state lead to broader spiral
features than in the standard case (Fig \ref{f:planetpos})}  
\label{f:planetposhigh}
\end{figure}

An ensemble of twelve  planetesimals was then introduced between $10.8$ and $18.8$ 
AU, aligned in four equispaced radial spokes (see  Figure  \ref{f:planetpos} for the 0.1 $M_*$ disc and  Figure  \ref{f:planetposhigh} for the 0.5 $M_*$ disc). The default
velocities were chosen to be Keplerian with a small correction for the 
gravitational potential of the disc, although we also investigated the case 
where 
the planetesimals were introduced with a non-zero initial eccentricity.

\section{Results}\label{s:results}

\subsection{Orbital evolution in the 0.1 solar mass disc case} 
\label{s:0.1 disc}
\begin{figure*}
\centering
  \includegraphics[width = 0.45\textwidth]{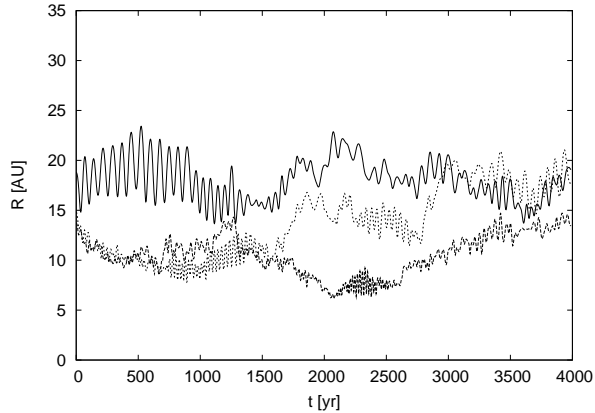}
  \caption[distance to te star]{Evolution of the distance to the star
of three representative planetesimals for
the $M_{\rm disc}  = 0.1 M_\odot$ case.
While individual planetesimals might migrate significantly over the course of the simulation, the average orbital radius only varies by less than 10\%.}
 \label{f:alla}
\end{figure*}

\begin{figure}
  \centering
  \includegraphics[width = 0.45\textwidth]{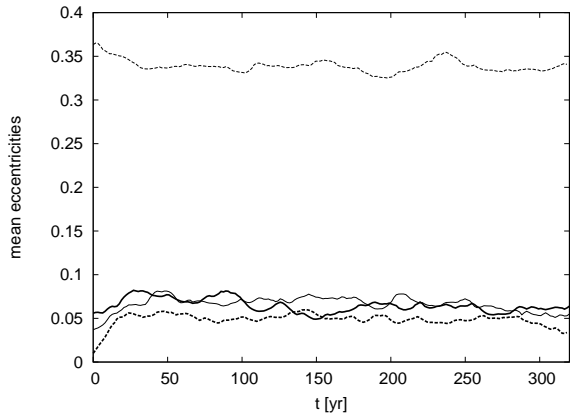}
  \caption[Mean eccentricity axis]{
Time evolution of the mean eccentricity of the planetesimals for
various initial eccentricities: 
$e(t=0)\approx 0$ (bold dashed line), $e(t=0)=0.06$ (bold solid line),
$e(t=0)=0.03$ (thin solid line) $e(t=0)=0.37$ (thin dashed line). During the course of the simulations a quasi-steady mean eccentricity $e\approx 0.07$ is established in all cases where the initial eccentricity is less than $0.1$.}
  \label{f:e_mean_vgl}
\end{figure}

In Fig. \ref{f:alla} we show the time evolution of the orbital radii
of all the planetesimals for the case where $M_{disc}=0.1M_{\star}$ and where the planetesimals are
initially set in nearly circular orbits.
Although some planetesimals undergo large amplitude excursions
(factor two or more), the average orbital radius of the ensemble varies
by no more than $10 \%$ during the simulation, indicating that planetesimals 
in a given region are apparently as likely to go inwards as outwards. 
The thick dashed  line in fig. \ref{f:e_mean_vgl} shows the mean eccentricity of the
planetesimals as a function of time, 
showing that $e \sim 0.05$ is typical (although individual planetesimals may
temporarily attain still higher eccentricities, $\sim 0.3$). Thus although much of the
fine structure in Figure \ref{f:alla} is simply a consequence of elliptical
orbital motion,  it is also evident that in addition  some planetesimals undergo
abrupt and/or large scale changes in orbital radius (for example, one planetesimal migrates over
$20 $ AU in 4000 years, and another follows a period of relative orbital
quiescence by migrating $\sim 5$ AU in about $200$ years).
Such migration rates imply radial velocities which are at times a significant
fraction of the Keplerian speed.

 In Figure \ref{f:e_mean_vgl}, we also show the results of starting the planetesimal 
ensemble with non-zero initial eccentricity, and see that the results are
unchanged for initial eccentricity less than $\sim 0.1$. On the other hand,
if a larger  initial eccentricity is
employed, the mean eccentricity of the ensemble remains around this value for
the duration of the experiment.

\subsection{Orbital evolution in the 0.5 solar mass disc case}
\label{s:0.5 disc}
\begin{figure*}
\centering
  \includegraphics[width = 0.45\textwidth]{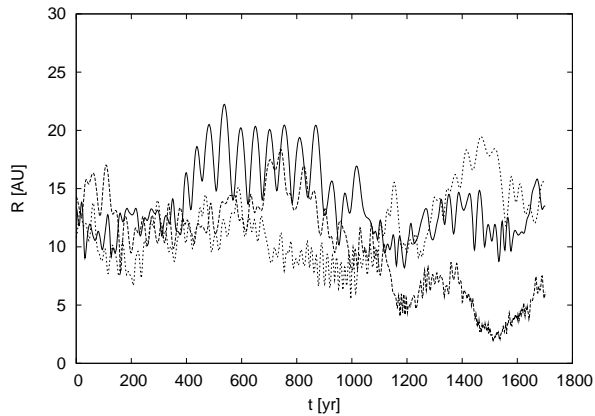}
  \caption[distance to te star]{Evolution of the distance to the star
of three representative planetesimals for
the $M_{\rm disc}  = 0.5 M_\odot$ case.}
 \label{f:alla_high}
\end{figure*}

 Figure \ref{f:alla_high} illustrates the rich variety of orbital histories
of planetesimals in the massive disc case. The amplitude of orbital migration
is somewhat larger than in the standard case, although again the change
in average orbital radius for the ensemble is small, $< 20 \% $. 
We note that, as in the standard case, the planetesimals undergo stochastic 
migration events even when they are located
at radii ($> 20 $ AU) where $Q$ is generally somewhat larger than unity.
These events coincide with spiral structures temporarily extending out into
these relatively quiescent regions.   

\begin{figure}
  \centering
  \includegraphics[width = 0.45\textwidth]{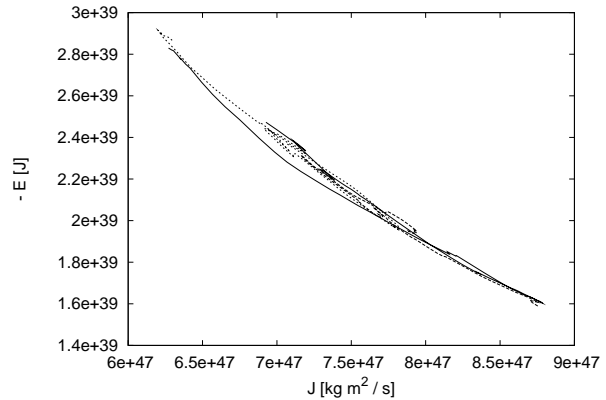}
  \caption[E versus L]{ Energy and angular momentum (in the inertial frame)
are plotted for three planetesimals in the massive disc case over the time
interval $550-700$ yr. The near linear relationship, with the
same slope, suggests that these three planetesimals are interacting with a mode
with pattern speed $\Omega_{\rm P} =4-5 \times 10^{-9}$ s$^{-1}$, i.e. with
co-rotation at $\sim 14$ A.U..}
\label{f:jacobi}\end{figure}

The mean eccentricity of the ensemble is about double its value in the standard case  
$ \sim 0.15$, with individual planetesimals temporarily
attaining eccentrcities of up to $0.5$. Although some of the quasi-periodic 
structure in Figure \ref{f:alla_high} is simply attributable to elliptical
orbital motion (i.e. at roughly constant angular momentum), 
we also see planetesimals in which, temporarily, the energy and angular momentum
are subject to correlated quasi-periodic variations. Such behaviour suggests
that these planetesimals are at this stage being driven by a  quasi-steady rotating
non-axisymmetric potential. We have investigated this possibility by noting
that in this case the Jacobi constant ($E_J$) of the planetesimal's motion (i.e the
total energy measured in the frame of the rotating pattern) 
should be conserved,  and that since $E_J$ is related to the energy, $E$,  in 
the inertial frame,  and angular momentum, $L$,  via

\begin{equation}
E_J= E-L \Omega_{\rm p}   
\end{equation}
(where $\Omega_{\rm p}$ is the pattern speed), we can test for such behaviour by
plotting $E$ versus $L$. This exercise indicates that there are indeed periods
when  certain planetesimals appear to be exhibiting this behaviour. This can be evidenced 
by a linear relation between $E$ and $L$, as shown in Figure \ref{f:jacobi}) for three of 
the simulated planetesimals. From the analysis of this relation, we can then determine the pattern
speed of the driving pattern. The planetesimals illustrated in Figure  \ref{f:jacobi}
appear to be interacting with a mode with co-rotation at around $14$ AU:
inspection of an animated series of snapshots of the simulation indicates that these 
three planetesimals are indeed roughly corotating with a spiral feature during this period, but that,
as the feature dissolves shortly thereafter, they then migrate and stall close to another spiral arm at a different radius. This behaviour is not seen in the standard ($0.1
M_\odot$ disc) simulations, which is consistent with the expectation (see, for example, \citealt{LR05}) that the more massive disc produces rather more coherent and
long-lived spiral features. Nevertheless, we find only occasional evidence
for planetesimals that are being driven by a dominant quasi-steady spiral mode so
that the analysis of orbital families in this situation (see e.g. \citealt{vorobyov06}) is of limited applicability here.


\subsection{Planetesimal-planetesimal encounters and gas  densities close to the planetesimals}

  Inspection of animations of the simulations suggest that planetesimals tend to
spend time preferentially in spiral structures in the disc.  As
such structures dissolve on a roughly dynamical timescale, the planetesimals then migrate
and find further temporary lodging in a new spiral structure. (The strongly
fluctuating nature of the spiral potential means that, as
discussed above,  the planetesimals usually have insufficient time to attain
steady orbits in the frame co-rotating with the local spiral pattern).
This predilection for being in the spiral arms suggests that the planetesimals
are expected to come closer to each other and to sample higher gas densities
than in the case of particles orbiting in an axisymmetric disc.

  We illustrate the increased tendency for planetesimal-planetesimal encounters
in Table \ref{t:encounters}, where we
contrast the number of encounters between the planetesimals in various
separation ranges with the corresponding statistics from a control run in 
which the planetesimals orbit in a smooth, non-self-gravitating disc. We see that the number of encounters with closest approach smaller than 0.5 AU is significantly enhanced in the self-gravitating disc case. As expected,
the effect of planetesimals lingering near spiral features is to decreasse the
minimum distance between planetesimals in the simulation. We stress that,
given the very low masses of the planetesimals (i.e. equal in mass to a single
gas particle), the effect of mutual encounters on the planetary
dynamics is negligible - as intended - even for the closest encounters
in the simulation. 

\begin{table}
  \centering
  \begin{tabular}{lcccc}
     & $d$ $<$ 0.5 & 0.5 $<$ $d$ $<$ 1 & 1 $<$ $d$ $<$ 1.5 & 1.5 $<$ $d$ $<$ 2 \\ 
    \hline
    0.1 $\msol$-disc & 10 & 17 & 17 & 19 \\
    non-self-grav. & 0 & 17 & 13 & 16 \\
  \end{tabular}
  \caption[Comparison of encounters]{Number of encounters with closest approach smaller than $d$ (measured in AU) between planetesimals during the first 15000 time units in the 0.1 $M_*$ disc run and the non-self-gravitating disc run.}
  \label{t:encounters}
\end{table}

\begin{figure}  \centering  
\includegraphics[width = 0.45\textwidth]{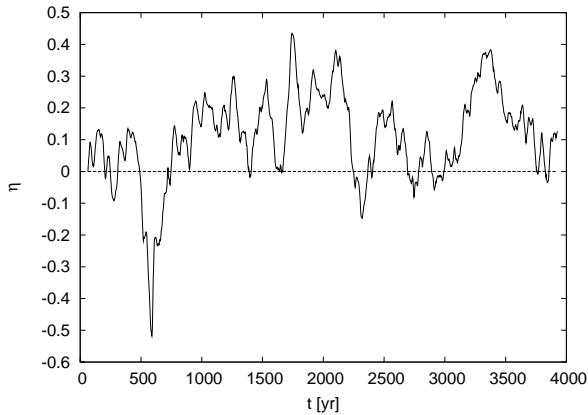}  
\caption[Density comparison for one planet]{Comparison between the density in the vicinity of one planetesimals in the 0.1 $M_*$ disc with the mean gas density at the current radius of the planetesimal.}
\label{f:rho_3}\end{figure}

 We illustrate the effect of  non-axisymmetric structure  in the disc on  the
gas density field sampled by the orbiting planetesimals in Fig. \ref{f:rho_3}.
We plot - for a given planetesimal - the density excess, which we define as 
\begin{equation}
\eta=\frac{2(\rho_{\rm p} - \rho_{\rm R})}{\rho_{\rm p} + \rho_{\rm R}},
\end{equation}
where $\rho_{\rm p}$ is the gas density in the vicinity of the planetesimal and 
$\rho_{\rm R}$ is the azimuthally averaged
density at the instantaneous orbital radius of the planetesimal.
This planetesimal appears to spend most of its time in the spiral arms and thus at larger gas densities. This is typical, although some planetesimals also undergo  phases where they sample  lower gas densities as can be seen in Fig. \ref{f:rho_10}.  Such periods in under-dense regions
are less apparent in the case of the  massive disc simulation.

  We see from Figure 7 that although this planetesimal samples gas densities 
larger than the local azimuthal average,
$\eta$ is always less than or of order unity.
This is to be expected:
the amplitude of the spiral features
in the gas is only of order unity and thus even if planetesimals spent {\it all}
their time in spiral arms, they would only experience a density that exceeded
the local mean by a modest factor.

\begin{figure}  \centering%
\includegraphics[width = 0.45\textwidth]{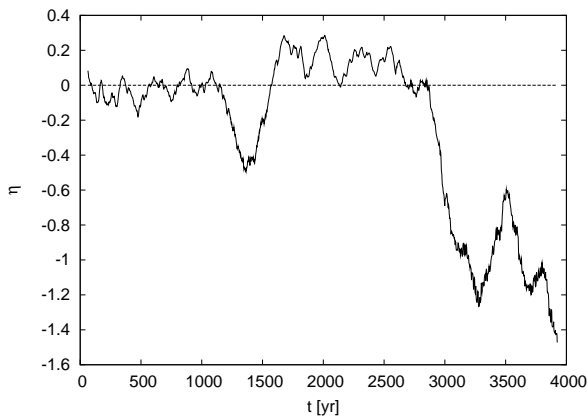}  
\caption[Density comparison for one planet]{
Evolution of the density excess  $\eta$ (see equation (3) as in Figure 7, but for a different planetesimal.}
\end{figure}

\subsection{Numerical issues}\label{s:numeric_concerns}

 Firstly, our experiments have aimed to study the response of each planetesimal
to the fluctuating disc potential and have only modeled the evolution of an 
ensemble of planetesimals for reasons of computational economy. Therefore, we 
need to be satisfied that the stochastic behaviour we see is a result of planetesimal-disc 
and not planetesimal-planetesimal interactions.  We find, in fact, that the torques experienced
by the planetesimals due to other planetesimals are typically two to three orders of
magnitude less than those arising from the disc. This is to be expected.
The closest planetesimal-planetesimal interactions occur at distances of $\sim 1 $ AU:
given that a typical smoothing length in the disc is $\sim 0.2$ AU and  that
a smoothing kernel contains $50$ particles, it follows that even at closest
approach there are about $6000$ gas particles that are closer to a planetesimal
than its nearest  planetesimal neighbour. Given that the planetesimal masses in the
simulations are equal to the gas particle masses, it follows that the
gravitational torques exerted between planetesimals are negligible.

 Secondly, we are interested here in examining the regime where the dominant torques experienced by the planetesimal derive from the non-axisymmetric density distribution that the disc would have {\it in the absence of the planetesimal} (i.e. that due to the spiral structure in the disk) rather than the non-axisymmetric structure induced in the disk by the planetesimal. Note
that in non-self gravitating discs, and thus in all discussions of
planetesimal migration in the literature (with the exception of that in
Nelson 2005) it is the latter effect that is responsible for planet
migration. In this case, it is essential that the calculation
properly resolves the Hill radius of the planetesimal 
\begin{equation}
R_h = R_{\rm p} \left(\frac{M_{\rm p}}{3 M_{\star}}\right)^{1/3},
\end{equation}
 where $R_{\rm p}$ is the radial coordinate of the planetesimal and $M_{\rm p}$ the mass of the planet,
 since this is the region from which the bulk of the torques on the planetesimal arise.

\begin{figure}  \centering%
\includegraphics[width = 0.45\textwidth]{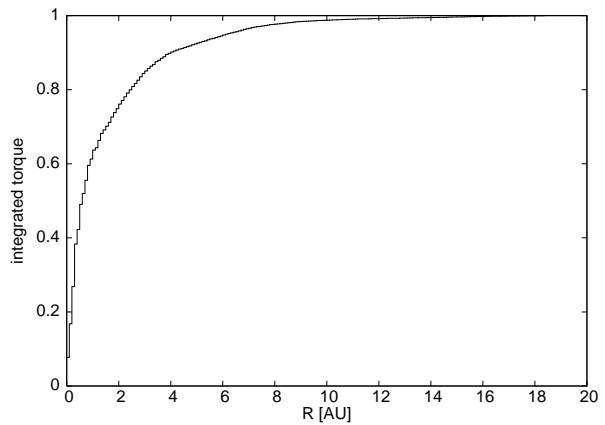}  
\caption[Example for the integrated torques acting on eccentricity]{Integrated 
z component of the torque  as a function of
distance from the planetesimal, normalised to the torque from within  20 AU
of the planetesimal.}
\label{f:tdense}\end{figure}

 In the present case, by contrast, where we examine migration driven
by spiral structure in the self-gravitating disc, it is no longer important
to resolve the Hill radius and we can thus examine the migration of low
mass planetesimals for which resolution of the Hill radius would be
impracticable. Nevertheless, we need to check that our assumption is correct,
i.e. that the dominant torques indeed arise at relatively large distances
from the  planet. Fig. \ref{f:tdense} illustrates a typical snapshot of the cumulative
distribution for the  z component of the 
torque of the disc on the planetesimal as a function of
distance from the planetesimal. We see that the dominant contribution to the torque 
is around 1 AU from the planetesimal and that the contribution
to the torque from the poorly resolved region around the planetesimal (\ie\
from within a typical particle smoothing length in that region, around 0.2 AU)
is small ($< 20 \%$). We are thus satisfied that the orbital evolution is not
driven by the disc's response to the planetesimal and that the planetesimal should
be behaving approximately  as a test particle. We examine this assumption
further by investigating how the migration depends on planetesimal mass
in Section \ref{s:validity}. 

On the other hand, in the case of 
the component of the torque perpendicular to z,
\ie\ acting on the inclination, the contributions from close to the planetesimal actually can make up a large fraction of the torque (i.e. around $40 \%$ of the torque originating
from within a smoothing length).  Thus the inclination development of the simulations cannot be trusted. Nevertheless the inclinations stay relatively small
(typically less than $0.02$ radians in the standard and $0.03$ for the
massive disc). This is small compared with the
axis ratio of the disc $H/R \sim 0.1$ and so the planetesimals remain
mainly confined within the disc.

\subsection{Valid regime}\label{s:validity}

 We now discuss the range of planetesimal masses for which the current
 calculation is valid. As noted in Section \ref{s:numerical}, we are
 modelling the regime in which the planetesimal behaves as a test
 particle in the gravitational potential of the star-disc system and
 where the results should be independent of planetesimal  mass. In reality, at low masses the effect of gas drag (omitted in the present simulations) would break this mass independence, whereas at high masses, the disc response to the presence of the planetesimal affects the torques on the planetesimal. We consider each of these issues in turn.

 A lower limit for the mass range of our simulation can be obtained by calculating the mass scale at which  the stopping time due to drag  is
comparable with  the 
timescale on which the planetesimal undergoes  stochastic migration
due to interaction with spiral structure in the disc. From inspection
of Figure \ref{f:alla} we see that planetesimals typically reverse their
direction of migration on timescales of about $10^3$ years. Thus if we
demand that the timescale for gas drag to significantly perturb the orbit
should exceed say $10^4$ years ($\sim 58$ orbits), we should safely be in the regime
where gas drag plays a minor role in the dynamics of the system. 

The drag force on solid bodies within a protostellar disc $F_{\rm D}$ is given by \citep{weiden77,whipple72}
\begin{equation}
F_{\rm D} = \frac{1}{2} C_{\rm D} \pi a^2 \rho u^2,
\end{equation}
where $a$ is the radius of the planetesimal, $u$ is the velocity of
the planetesimal relative to the gas, $\rho$ is the gas density  and $C_{\rm D}$ is the drag coefficient.



Thus the timescale on which relative motion between the planetesimal and the disc
gas is damped by gas drag is 
\begin{equation}
t_{\rm e} = \frac{M_{\rm p} u}{F_{\rm D}}.
\end{equation}
  In the case that we are considering here, the relative velocity between
the planetesimal and the disc gas does not derive - as in the case usually
considered  (e.g. Weidenschilling 1977) - from the difference between Keplerian planetesimal motion and
sub-Keplerian gas motion subject to outward pressure forces. Instead,
this relative velocity, $u$, derives from the eccentricity of the
planetesimal  orbits induced by the spiral structure in the disc, so
that we have $u = e V_{\rm K}$, where $V_{\rm K}$ is the local
Keplerian velocity. 

Assuming a spherical planetesimal of density $\rho_p$, we can then write $t_{\rm e}$
as

\begin{equation}
t_{\rm e} = \frac{8 \rho_p R}{ 3 C_D \rho e V_{\rm K}}
\end{equation}

or, equivalently (and adopting $C_D$ in the range $0.5-1$, which is appropriate
to the case considered here where the Reynolds number exceeds a few hundred)

\begin{equation}
t_{\rm e} \sim  \frac{8}{1.32}\frac{\rho_p}{\rho} \frac{a}{R} \frac{\Omega^{-1}}{e}
\end{equation}
where $R$ is the orbital radius of the planetesimal. If we then require that
$t_{\rm e} > 10^4$ years ($\sim 300 \Omega^{-1}$), and adopting typical
values of $ e \sim 0.1$, $\rho \sim $ a few $ \times 10^{-11}$ g cm$^{-3}$ and
$\rho_p \sim $ a few g cm$^{-3}$,  we then obtain
the requirement that $a$ must be at least of order a kilometre or so before one can
neglect the effect of gas drag. 

We thus conclude that the effect of gas drag is negligible for planetesimals of
kilometre scale or above (corresponding to bodies of masses $\sim 10^{16}$ g.)

\begin{figure}  \centering
\includegraphics[width = 0.45\textwidth]{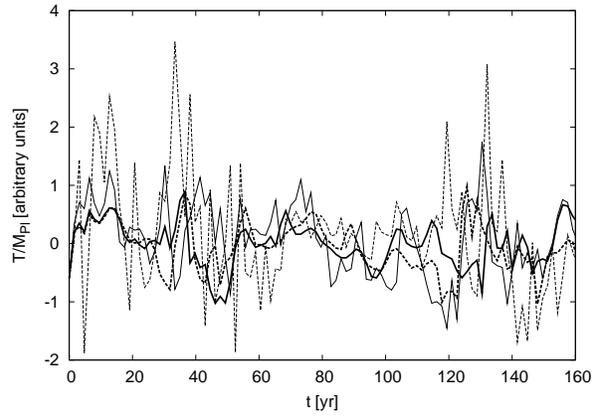}  
\caption[Comparison of torques in different runs]{The torque per unit
  planetesimal mass as a function of time
for runs in which the  mass of the planetesimal is respectively $1/10$
  (bold dashed line), $1$ (bold solid line), $10$ (thin solid line)
and $100$ (thin dashed line) times the  mass of a  gas particle.}
\label{f:tenth}
\end{figure}

\begin{figure}
  \centering
  \includegraphics[width = 0.45\textwidth]{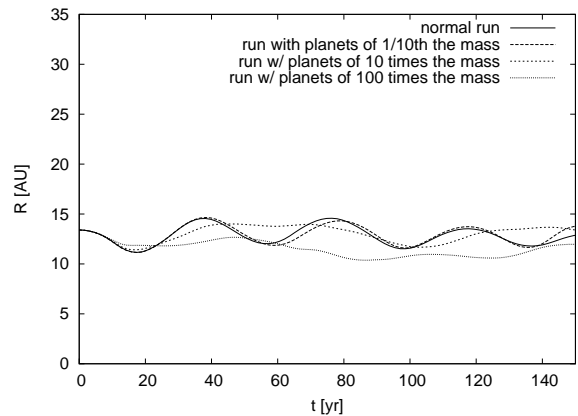}
  \caption[Distance to the star for different mass planets]{
The distance to the star is plotted against time for one planetesimal in the normal run (solid line), the run with a tenth planetesimal mass (long dashed line), the run with 10 times higher planetesimal masses (short dashed line) and the run with 100 times higher planetesimal masses (very short dashed line).}
  \label{f:comp_a}
\end{figure}

Our simulations are valid for all higher masses as long as the accelerations 
experienced by the planetesimals remain independent of mass. But these accelerations
change as soon as the planetesimals have enough mass to gravitationally influence the gas density. Thus we get an upper limit for the validity of our simulations by checking at what mass scale  the torque per planetesimal mass $T/M_{\rm p}$ changes with mass.
In Figure 10 we plot the evolution of $T/M_{\rm p}$ for a single planetesimal
in the case that it has the mass indicated in each of the curves.
It can be seen in Fig. \ref{f:tenth}  that significant changes to the torques on the planetesimals only arises when the planetesimal  mass is raised above ten times the one used in our standard run, corresponding to about 
$8 \times 10^{27} \gram$. We reach the same conclusion from inspection
of Fig. \ref{f:comp_a}, where again we see a marked deviation in orbital
evolution for the largest mass planetesimal.

So summarising, we have as a region of validity of our simulations:
\begin{equation}
2 \ex{-8} M_{\oplus} \sim   10^{16} \gram < M_{\rm p} < 8\ex{27} \gram = 1 M_{\oplus}
\end{equation}
or in planetesimal  radii
\begin{equation}
1 {\rm km} < a < 5000 {\rm km}. 
\label{e:valid}
\end{equation}

\section{Discussion}
We can summarise our numerical findings as follows. Planetesimals
in the size/mass range to which our  calculations apply (see equation
\ref{e:valid}) are subject to large scale fluctuations in orbital
radius due to interaction with regenerative spiral structure in the disc.
We can discern no net sign to the mean migration rate over the
timescale of the simulation and are thus unable to answer one of the goals
of the present study, i.e. whether we expect planetesimals to migrate
into the star during the self-gravitating phase of the disc. Since
the disc's lifetime in the self-gravitating phase exceeds the duration
of the present experiment by about a factor $30$ it will be computationally
challenging to answer this question, especially since (given the
stochastic nature of the orbital evolution) it will be necessary to
assess the fraction of planetesimals retained in the disc through
following  the evolution of a large ensemble of particles.

  We can however make some comments about how the dynamical evolution
of our particles is likely to affect their collisonal growth during the
self-gravitating phase.  As noted by Nelson (2005) in the context
of the evolution of planetesimals in a disc subject to MRI, the fluctuations
in orbital radius and eccentricity could in principle either aid or
hinder planetesimal growth. On the positive side, orbital migration provides
a mechanism for avoiding the self-limitation of planetesimal growth
through the clearing out of a proto-planet's ``feeding zone". In
conventional models, this imposes an ``isolation mass" limit which makes
it difficult to grow beyond a  terrestrial mass scale in the inner
regions of the disc \citep{ida04}. Although this limitation is by-passed in
our self-gravitating models, since the feeding zone would be continuously
replenished by planetesimals undergoing an orbital random walk, this
factor is not very significant since in any case the high
surface density during the self-gravitating phase means that isolation masses
are high (higher, that is, than the upper limit for the applicability
of our calculations). Another potentially positive aspect of evolution
in a self-gravitating disc is the possibility of concentrating planetesimals
in high density spiral features, where they might be more prone to
undergo collisional growth. In the case of the more massive -- $0.5 M_\odot$ --
disc (see Section 4), there is some evidence for the dominance
at times of spiral modes which persist over a number of dynamical times,
allowing particles to settle into orbits which conserve a Jacobi integral
in the co-rotating frame. Some of these orbits involve a preference for
particles to spend time in the spiral arms, an impression strengthened
by inspection of animations which show particles which co-rotate for
a while in the vicinity of an  arm and which then respond to changes
in the spiral structure by passing  rapidly to another
spiral feature. This tendency will not lead to a significantly enhanced
collision rate, however, unless there is a strong enhancement in the
{\it solid to gas ratio} in the spiral arms: since the amplitude of the
gaseous arms is around a factor two at most, then even a particle
that spent {\it all} its time in the arms would only, at fixed solid to dust
ratio, experience an enhancement in the local planetesimal density
of a factor two at most. Since we have only studied the evolution
of ten planetesimals, we cannot, on the basis of the present study,
comment on the degree to which planetesimals are differentially
concentrated in the arms. Nevertheless, the results of \citet{rice04}
suggest that the differential accumulation of solids in the arms
is weak at  size scales where gas drag is unimportant.

  We now turn to the negative implication of stochastic migration
 noted by Nelson (2005), i.e. the growth of planetesimal velocity
dispersion. This hinders planetesimal growth in two respects: higher
collisional velocities are associated with disruption of parent bodies
instead of collisional growth (e.g. Leinhardt et al 2007) and also
suppress the role of  gravitational focusing in effecting physical collisions.
 Quantifying the second effect, the timescale for collisional growth
of a planetesimal of mass $M_{\rm p}$, density $\rho$ and  radius
$R_{\rm p}$ is given by:
\begin{equation}
t_{\rm grow} =  \frac{ \rho^{2/3} M_{\rm p}^{1/3}}{\Sigma_p \Omega (1+4GM_{\rm p}/(R_{\rm p} \sigma^2))}
\end{equation}
where $\Sigma_{\rm p}$ is the surface density of planetesimals, $\sigma$
is the planetesimal velocity dispersion and
$\Omega$ is the local Keplerian orbital frequency.
 For a planetesimal of density $\sim 3$
g/cm$^3$  located at $10$ A.U. in a disc with surface
density of planetesimals  equal to 0.4 g/cm$^ {2}$, 
the growth timescale via two body 
collisions is:
\begin{equation}
t_{\rm grow} = 5\times 10^9 \frac{\tilde M^{1/3}}{1+\tilde M^{2/3}\tilde\sigma^{-2}}~~\mbox{yrs}
\end{equation}
where $\tilde M=(M_p/6 ~10^{-4}M_{\oplus})$ 
and $\tilde \sigma=(\sigma/1 \mbox{km s}^{-1})$. Evidently this timescale
is far too long to be of interest unless the second term on the
denominator (due to enhancement of the collisional cross section by
gravitational focusing) is large. {\it If} we simply set $\sigma$ to be the
product of the orbital eccentricity and local Keplerian velocity (an
assumption we revisit below),  it follows that at a given location
in the disc and for a given mass scale of planetesimal, 
then the growth
timescale in the gravitationally focused regime scales 
as the {\it square of the eccentricity} divided by the
surface density normalisation. In other words,
for a planetesimal at a given mass scale to be able to grow significantly
over a disc phase of duration $t_{\rm life}$ we require
\begin{equation}
 \frac  {t_{\rm grow}} {t_{\rm life}} < 1
\end{equation}
which translates into a lower limit on the 
product $t_{\rm life} \Sigma e^{-2}$
(where $\Sigma$ is the surface density normalisation of the disc).

 We can now return to the issue raised in the Introduction of whether or not
planetesimal growth is favoured in the high density but short lived
phase during which the disc is self-gravitating. Assuming for simplicity
that the self-gravitating phase lasts $\sim 10 \%$ of the subsequent
lifetime of the primordial gas/dust disc but that its surface density
(mass) is typically ten times greater, we see that these two factors
roughly cancel. Thus whether the growth of planetesimals is favoured
during the self-gravitating phase actually boils down to the relative
values of the eccentricity in the two phases.

  Our simulations have shown that this factor is decisive in disfavouring
the self-gravitating regime. In conventional models for the dynamical
evolution of planetesimal swarms (e.g. \citealt{kokubo2000}) the equilibrium
eccentricity (attained through the balance between stirring by larger
bodies and damping by gas drag) is very low (of order $0.01$). We have
however seen that in the self-gravitating phase, the typical eccentricity
is $0.1$ or more. Since the growth time scales as the square of the
eccentricity, this means that the self-gravitating phase is highly
unfavourable (compared with the later more quiescent phase) for the
collisional growth of planetesimals. We note that the above
estimates for conventional proto-planetary discs assume a laminar
disc structure;  eccentricity driving by interaction with
turbulent structures in a disc exhibiting MRI would further hinder
collisional growth, whether or not the disc is self-gravitating.

  There however remain two possible qualifiers of the above conclusion.
Firstly,  it is only valid to set the local velocity dispersion
equal to the product of the eccentricity and local Keplerian velocity
in the case that the kinematics is well described as a set of elliptical
orbits which are elongated in a random direction. Such a description
does not of course allow for the possibility of {\it local velocity
coherence} in a planetesimal swarm in which velocities are significantly
non-circular. In this sense, therefore, the above estimates
are pessimistic. Further calculations, involving a large swarm of
planetesimals from which one could derive the {\it local} velocity
dispersion, are required in order to settle this issue. We however note
that our difficulty in identifying coherent orbital families, interacting
with a long lived spiral mode, probably implies that this effect is unlikely 
to improve the prospects for particle growth significantly. The second
possible qualifier is that, as noted in Section 2, our simulations employ
a cooling timescale which is close to the critical one for disc fragmentation
and therefore represent conditions where spiral modes are of relatively
high amplitude. In the inner regions of protoplanetary discs, however, cooling
timescales are longer than employed here and thus one would expect lower 
amplitude spiral structures, with a correspondingly lower amplitude of
stochastic driving of planetesimal orbits. This possibility needs to
be quantified through further numerical investigation.

  If one leaves aside these possibilities, then we have shown that the
self-gravitating phase is unfavourable for collisional growth in the regime
of planetesimal scale studied. We have not ruled out the possibility
of collisional growth at smaller scales (where gas drag can effect
a significant enhancement of the local solid to gas ratio in spiral
features; \citealt{rice04}) nor that (where such enhancement exists)
it may not be possible to produce much larger bodies through
the gravitational fragmentation of the solid component (e.g., \citealt{rice06}).
We nevertheless consider it likely that whatever the size scale
attained in this way, there will be no further collisional growth
of planetesimals while the disc is self-gravitating.

  This obviously limits the extent to which planet formation can get under way
in the self-gravitating phase. Either a massive planet is produced
via gas phase Jeans instability \citep{boss00}, thus obviating the
need for collisional growth, or else, if self-gravity aids the
initial accumulation of solid bodies \citep{rice04,rice06},  then it is
unlikely that such bodies can undergo further collisional growth once
they become large enough for gas drag to be unimportant. In this latter
case, further growth of such bodies could only resume once the disc
had evolved to the point where its self-gravity (and the associated
pumping up of the planetesimal velocity dispersion) has become unimportant.
The vital unanswered question, therefore, is whether orbital migration
would prevent the planetesimals from remaining in the disc
throughout the self-gravitating phase. This question can only be
answered through further, computationally challenging, calculations.

\section{Acknowledgments}
M. Britsch gratefully acknowledges the hospitality of the IOA and
financial support via the  Marie Curie EARA Early
Stage Training program. We thank the referee, Phil Armitage, for
constructive criticism.
\bibliographystyle{mn2e} 
\bibliography{lodato}

\end{document}